\documentclass[prd,twocolumn,nofootinbib,reprint]{revtex4-1}
\usepackage{graphicx}
\usepackage{cancel}
\usepackage{amssymb}
\usepackage{textcomp}
\usepackage{amsmath}
\usepackage{bm}
\usepackage{times}
\usepackage{epsfig}
\usepackage{color}
\usepackage{graphics}
\usepackage{hyperref}
\usepackage{setspace}
\usepackage{slashed}

\hypersetup{
    pdfnewwindow=true,      % links in new window
    colorlinks=true,       % false: boxed links; true: colored links
    linkcolor=black,          % color of internal links
    citecolor=blue,        % color of links to bibliography
    filecolor=blue,      % color of file links
    urlcolor=blue           % color of external links
}

\def\bea{\begin{eqnarray}}
\def\eea{\end{eqnarray}}
\begin{document}
\title{Gravitational Waves as a New Probe of Bose-Einstein Condensate Dark Matter}
\author{P. S. Bhupal Dev$^{1,2}$, Manfred Lindner$^1$, and Sebastian Ohmer$^1$}
\address{$^1$Max-Planck-Institut f\"{u}r Kernphysik, Saupfercheckweg 1, D-69117 Heidelberg, Germany}
\address{$^2$Department of Physics and McDonnell Center for the Space Sciences,
Washington University, St. Louis, MO 63130, USA}
%%%%%%%%%%%%%%%%%%%%%%%%%%%%%%%%%%%%%%%%%%%%%%%%%%%%%%%%%%%%%%%%%%%%
%\date{\today}
%%%%%%%%%%%%%%%%%%%%%%%%%%%%%%%%%%%%%%%%%%%%%%%%%%%%%%%%%%%%%%%%%%%%
%\vspace{1.0cm}
\begin{abstract}
There exists a class of ultralight Dark Matter (DM) models which could form a Bose-Einstein condensate (BEC) in the early universe and behave as a single coherent wave instead of individual particles in galaxies. We show that a generic BEC-DM halo intervening along the line of sight of a gravitational wave (GW) signal could induce an observable change in the speed of GWs, with the effective refractive index depending only on the mass and self-interaction of the constituent DM particles and the GW frequency. Hence, we propose to use the  deviation in the speed of GWs as a new probe of the BEC-DM parameter space. With a multi-messenger approach to GW astronomy and/or with extended sensitivity to lower GW frequencies, the entire BEC-DM parameter space can be effectively probed by our new method  in the near future.
\end{abstract}
%%%%%%%%%%%%%%%%%%%%%%%%%%%%%%%%%%%%%%%%%%%%%%%%%%%%%%%%%%%%%%%%%%%%
\maketitle
%%%%%%%%%%%%%%%%%%%%%%%
\section{Introduction} \label{sec:1} 
%%%%%%%%%%%%%%%%%%%
Although the existence of Dark Matter (DM) constituting about 27\% of the energy budget of our Universe~\cite{Ade:2015xua} is by now well established through various cosmological and astrophysical observations, very little is known about its particle nature and interactions. While the standard $\Lambda$CDM model with collisionless cold DM (CDM) successfully explains the large-scale structure formation by the hierarchical clustering of DM fluctuations~\cite{Blumenthal:1984bp, Davis:1985rj}, there are some unresolved issues on galactic and sub-galactic scales, such as the core-cusp~\cite{Moore:1994yx, Flores:1994gz, Moore:1999gc, deBlok:2009sp}, missing satellite~\cite{Kauffmann:1993gv, Klypin:1999uc, Moore:1999nt, Bullock:2010uy}, and too big to fail~\cite{BoylanKolchin:2011de, BoylanKolchin:2011dk, Papastergis:2014aba} problems. All these small-scale structure anomalies can in principle be resolved if the DM is made up of ultralight bosons that form a Bose-Einstein condensate (BEC), i.e. a single coherent macroscopic wave function with long range correlation; for a review, see e.g.,~Ref.~\cite{Suarez:2013iw}. 

There are two classes of BEC-DM, depending on whether DM self interactions are present or not. Without any self interactions, the quantum pressure of localized particles is sufficient to stabilize the DM halo against gravitational collapse only for a very light DM with mass $m\sim 10^{-22}$ eV~\cite{Sahni:1999qe, Hu:2000ke, Sin:1992bg, Matos:2008ag, Lee:2008jp, Lora:2011yc}, whereas a small repulsive self-interaction can allow a much wider range of DM masses up to $m\lesssim 1$ eV~\cite{Peebles:2000yy, Goodman:2000tg, Arbey:2003sj, Eby:2015hsq, Fan:2016rda}.\footnote{BEC configurations with heavier DM and/or an attractive self-interaction are usually unstable against gravity~\cite{Khlopov:1985jw} and more likely to form local dense clumps such as Bose stars~\cite{Tkachev:1986tr, Colpi:1986ye, Tkachev:1991ka, Kolb:1993zz, Lee:1995af}, unless the thermalization rate is faster than the Hubble rate to overcome the Jeans instability.} Concrete particle physics examples for BEC-DM are WISPs (Weakly Interacting Slim Particles)~\cite{Ringwald:2012hr}, which include the QCD axion or axion-like particles~\cite{Sikivie:2009qn, Mielke:2009zza, Erken:2011dz, Saikawa:2012uk, Davidson:2013aba, Berges:2014xea, Davidson:2014hfa, Guth:2014hsa, Banik:2015sma} and hidden-sector gauge bosons~\cite{Nelson:2011sf, Arias:2012az, Pires:2012yr, Soni:2016gzf} ubiquitous in string theories, but our subsequent discussion will be generically applicable to any BEC-DM with a repulsive self-interaction, which is necessary to obtain long-range effects~\cite{Guth:2014hsa}.\footnote{Although the simplest models,  where the scalar potential has an approximate symmetry to ensure the radiative stability of the ultralight scalar, usually give rise to an attractive self-interaction in the non-relativistic limit, it is possible to have realistic models with repulsive self-interaction~\cite{Fan:2016rda, Berezhiani:2015bqa}.}
 
The observational consequences on structure formation mentioned above cannot distinguish a BEC-DM from an ordinary self-interacting DM~\cite{Spergel:1999mh}. Existing distinction methods include enhanced integrated Sachs-Wolfe effect~\cite{Sikivie:2009qn}, tidal torquing of galactic halos~\cite{Banik:2015sma, RindlerDaller:2011kx, Banik:2013rxa}, and effects on cosmic microwave background matter power spectrum~\cite{Ferrer:2004xj, Velten:2011ab}. We propose a new method to probe the BEC-DM parameter space using gravitational wave (GW) astronomy, inspired by the recent discovery of transient GW signals at LIGO~\cite{Abbott:2016blz, Abbott:2016nmj}. We show that if GWs pass through a BEC-DM halo on their way to Earth, the small spacetime distortions associated with them could produce phononic excitations in the BEC medium which in turn induce a small but potentially observable change in the speed of GWs, while the speed of light remains unchanged. This approach is very effective if any of the future multi-messenger searches  for gamma-ray, optical, $X$-ray, or neutrino counterparts to the GW signal become successful. On the contrary, a lack of any observable deviation in the speed of GWs will put stringent constraints on the BEC-DM scenario. In fact, we find that even with the current LIGO sensitivity, it might be possible to partly rule out the BEC-DM parameter space otherwise preferred by existing cosmological data. Future GW detectors such as eLISA~\cite{Seoane:2013qna} with extended sensitivity to lower GW frequencies will be able to completely rule out the cosmologically preferred region.

The rest of the paper is organized as follows: in Section~\ref{sec:2}, we calculate the change in the speed of GWs due to energy loss inside the BEC medium. In Section~\ref{sec:3}, we apply this result to derive constraints on the BEC-DM parameter space. In Section~\ref{sec:4}, we discuss the effect of gravitational lensing. Our conclusions are given in Section~\ref{sec:5}. 
%%%%%%%%%%%%%%%%%%%%%%%%%%%%%%%%%%%%%%%%%%%%
\section{Speed of GW inside BEC medium} \label{sec:2} 
The cosmological dynamics of BEC-DM can be described by a single classical scalar field $\phi$~\cite{Hertzberg:2016tal}, with the effective Lagrangian 
 \begin{align}
\mathcal{L} \ = \ \frac{1}{2} \partial_\mu \phi \partial^\mu \phi - \frac{1}{2} m^2\phi^2 - \lambda \phi^4  \, ,
\label{eq:lag}
\end{align}
analogous to the Ginzburg-Landau free energy density in a neutral superfluid. A real scalar field will suffice for our discussion.  
In~\eqref{eq:lag}, we have considered a simple renormalizable scalar potential with only quadratic and quartic terms, the latter providing a repulsive self-interaction for the DM, as required in addition to the quantum pressure of localized particles to stabilize the DM halo core against gravitational collapse. For no self-interaction ($\lambda=0$), the quantum pressure is sufficient only if $m\sim 10^{-22}$ eV, a scenario known as fuzzy DM~\cite{Hu:2000ke}. In principle, we could also have added a cubic term $-g \phi^3$ to~\eqref{eq:lag}; however, for the self-interaction to be repulsive in the non-relativistic limit, we must have $\lambda> 5g^2/2$~\cite{Fan:2016rda}. Similarly, we do not include any higher-dimensional operators in~\eqref{eq:lag}. 

Using~\eqref{eq:lag}, we calculate the stress-energy tensor 
\begin{align}
T^{\mu\nu} \ = \ \frac{\partial {\cal L}}{\partial(\partial _\mu \phi)}\partial^\nu \phi - g^{\mu\nu}{\cal L} \, ,
\label{eq:stress}
\end{align}
where $g^{\mu\nu}$ is the spacetime metric. Far from the GW source, the linearized spacetime metric can be written as $g^{\mu\nu}= \eta^{\mu\nu}+h^{\mu\nu}$, where $\eta ={\rm diag}(1,-1,-1,-1)$ is the flat Minkowski metric (in particle physics conventions) and $h^{\mu\nu}$ is a small perturbation. To leading order, the background mean field values of the energy density $\rho_0 \equiv T^{00}$ and pressure $p_0 \equiv T^{ii}$ of the BEC medium are related by the equation of state (EoS)
\begin{align}
p_0 \ = \ \frac{3}{2}\frac{\lambda}{m^4}\rho_0^2 \, .
\label{eq:eos1}
\end{align}

Gravity is a long-range force. Because almost all particles in the BEC system are condensed into the lowest energy available state with very long de Broglie wavelength, the GWs can excite the massless phonon modes in the ground state of the BEC wave function~\cite{Sabin:2014bua}. As a result, the GW undergoes enhanced coherent forward scattering inside a BEC-DM halo compared to an ordinary CDM halo. This effect is analogous to light traveling through an optically transparent medium (e.g. glass) with refractive index different from $1$.  In general, there could be either refraction or absorption of the incident wave (apart from reflection), depending on the real or imaginary part of the refractive index, respectively. The refraction effect modifies the wavenumber and propagation speed of the wave in the medium (without change in its frequency and amplitude), while absorption results in the damping of the amplitude, and hence, attenuation of the wave in the medium.  In the case of GWs incident on a BEC medium, the absorption effect is negligible, because it would require exciting the phonons from massless to massive modes, which in turn requires much larger energy comparable to the chemical potential of the BEC~\cite{Sabin:2014bua}. Thus, only the propagation speed of the GW passing through a BEC-DM halo is reduced, but much more strongly than in a CDM halo, as we show below.

To estimate this effect, we first write down the effective metric of the BEC phononic excitations on the flat spacetime metric~\cite{Sabin:2014bua, Fagnocchi:2010sn, Visser:2010xv}
\begin{align}
g_{\rm eff} \ = \ \frac{n_0^2}{c_s(\rho_0+p_0)}{\rm diag}(c_s^2, -1, -1, -1) \, ,
\label{eq:geff}
\end{align}
where $n_0 \equiv \rho_0/m$ is the number density of the background mean field and 
\begin{align}
c_s \ \equiv \ \left(\frac{\partial p_0}{\partial \rho_0}\right)^{1/2} \ = \ \left(\frac{3 \lambda \rho_0}{m^4}\right)^{1/2}
\end{align}
is the speed of sound obtained from the background EoS~\eqref{eq:eos1}. The solution to the Klein-Gordon equation with the metric in~\eqref{eq:geff} thus describes massless excitations propagating with the speed $c_s$. The frequency of the mode satisfies the linear dispersion relation $\omega_k=c_s|\mathbf{k}|$, where $\mathbf{k}$ is the 3-momentum  of the mode.

The refractive index of a GW scattering off a gravitational potential was first calculated in~\cite{Peters:1974gj} and was shown to be negligibly small for ordinary matter. Here, we provide an alternative derivation of the refractive index based on the optical theorem and argue that it could be relevant in our case due to the enhanced forward scattering rate in a BEC medium. The optical theorem links the refractive index $n_g$ to the forward scattering amplitude $f(0)$ of the incident wave with the scatterers inside the medium:
\begin{align}
n_g = 1 + \frac{2 \pi n f(0)}{k^2}\,,
\end{align}
with $n$ the number density of scatterers inside the medium and $k$ the wavenumber of the incident wave. We estimate the forward scattering $n f(0)$ of the GW in the BEC-DM halo by relating the energy density of the incident GW to the energy density of the massless phonon excitations in the ground state.

We assume for simplicity that the phonons can be described by a one-dimensional wave function with hard-wall boundary conditions. This approximation is also valid for a  spherically-symmetric DM halo, such that only the radial component matters for the GW propagation through it. 
The energy spectrum of the massless modes is then given by
\begin{align}
\omega_l \ = \ \frac{l \pi c_s}{\langle D_{\rm halo}\rangle} \,,
\label{eq:6}
\end{align}
where $\langle D_{\rm halo}\rangle = 4R/\pi$ is the average distance the gravitational wave propagates through the spherically-symmetric DM halo with a radius $R$ and $l\in \{1,2,\cdots\}$. Therefore, the {\it minimum} energy density required to excite the massless phonon modes in the BEC medium is given by
\begin{align}
\Delta \rho \ \equiv \ n \Delta \omega \ = \ \frac{n\pi^2 c_s}{4 R}\,, 
\label{eq:7}
\end{align}
where $n$ is the number density of phonons in the BEC and $\Delta \omega\equiv \omega_{l+1}-\omega_l$ is the energy difference between two adjacent massless modes [cf.~\eqref{eq:6}].

The radius $R$ of the gravitationally bound BEC with a repulsive self-interaction only depends on the physical characteristics of the particles in the condensate~\cite{Boehmer:2007um, Chavanis:2011zi, Chavanis:2011zm}:
\begin{align}
R \ = \ \left(\frac{\pi^2 a_s}{Gm^3}\right)^{1/2} ,
\label{eq:R}
\end{align}
where $a_s$ is the $s$-wave scattering length which in the low-energy limit is defined by $\displaystyle\lim_{k\to 0}\sigma(\phi \phi \to \phi \phi) = 4\pi a_s^2$. For the interaction Lagrangian given by~\eqref{eq:lag}, we find
\begin{align}
\sigma \ = \ \frac{9\lambda^2}{\pi m^2} \, ,
\label{eq:sigma}
\end{align}
and hence, from~\eqref{eq:R}, we can extract 
\begin{align}
R \ = \ 2\pi \sqrt{3\lambda}\: \frac{M_{\rm Pl}}{m^2} \, .
\label{eq:R2}
\end{align}

The number density of phonons in the BEC can be estimated in terms of the microscopic parameters from the expression for the critical temperature~\cite{Bettoni:2013zma}
\begin{align}
T_c \ = \ \frac{2\pi}{m}\left( \frac{n}{\zeta(3/2)} \right)^\frac{2}{3} \,,
\end{align}
and equating it to the critical temperature for a self-interacting scalar gas: $T_c  =  (24m^2/\lambda)^{1/2}$~\cite{Dolan:1973qd, Weinberg:1974hy, Kapusta:1981aa}. We thus obtain 
\begin{align}
n \ = \ \left( \frac{6 m^4}{\pi^2 \lambda}\right)^\frac{3}{4} \zeta(3/2)\,.
\end{align}

The typical energy density of a GW is given by~\cite{Misner:1974qy}
\begin{align}
\rho_{\rm{GW}} \ = \ \frac{1}{4}M^2_{\rm Pl}\omega_{\rm GW}^2 h^2 \,,
\end{align}
where $\omega_{\rm GW}=2\pi f$ is the angular frequency ($f$ being the frequency) of the GW and $h$ is the amplitude (we assume $h_+ = h_\times = h/\sqrt{2}$ for the two polarization modes), and $M_{\rm Pl}$ is the Planck mass. Using a linear dispersion relation, we can calculate the relative change in the wavenumber of the GW due to its propagation in the BEC 
\begin{align}
\frac{\Delta \rho}{\rho_{\rm{GW}}} \ = \ 2\frac{\Delta k}{\omega_{\rm GW}}
\label{eq:14}
 \,,
\end{align}
where $\Delta \rho$ is given by~\eqref{eq:7} and $\Delta k$ is the change in the wavenumber of GWs (which can be thought of as the induced mass of the graviton, if the GWs were quantized). Thus, the effective refractive index is given by 
\begin{align}
n_g \ =\ 1+\frac{\Delta k^2}{2\omega_{\rm GW}^2} \, .
\label{eq:ng}
\end{align}
%{\color{yellow} which can also be understood as a second order effect arising from the interaction of the scalar
%gravitational potential and the GWs~\cite{Peters:1974gj}.} 
This effect is negligible for ordinary CDM (or for ordinary matter per se), since the number density in any given energy eigenstate is too small, which makes  $\Delta k$ unobservable~\cite{Elghozi:2016wzb}. However, the huge occupation number in the ground state and the long-range correlations of the BEC system could enhance this effect sizably, and we will exploit this key feature to derive constraints on the BEC-DM parameter space.

Thus, the change in refractive index experienced by the GWs inside the BEC medium is given by 
\begin{align}
\delta n_g \ \equiv \  n_g-1 \ = \ \sqrt{\frac{3}{2}} \frac{3 m^6 \rho_0 \zeta(3/2)^2}{8 \pi \lambda^{3/2} h^4 \omega_{\rm GW}^4 M^6_{\rm{Pl}} } \,,
\label{eq:dng}
\end{align}
which depends on the average DM density $\rho_0$, the amplitude $h$ and angular frequency $\omega_{\rm GW}$ of the GW and the microscopic parameters $(m,\lambda)$. Fixing the macroscopic parameters $(\rho_0, h, \omega)$ we can directly translate a constraint on the speed of GWs into a constraint on the $(m,\lambda)$ parameter space of BEC-DM, as shown in Section~\ref{sec:3}.

The density distribution of a static, spherically-symmetric BEC-DM halo can be obtained from the solution to the Lane-Emden equation~\cite{chandra} in the weak field, Thomas-Fermi regime, given by the analytic form~\cite{Boehmer:2007um}   
\begin{align}
\rho(r) \ = \ \rho_{\rm cr} \frac{\sin{\kappa r}}{\kappa r} \, ,
\label{eq:halo}
\end{align}
where $\kappa=\sqrt{Gm^3/a_s}=\pi/R$ and $\rho_{\rm cr}$ is the central density of the condensate. The average density of a BEC-DM halo is thus given by 
\begin{align}
\rho_0 \ \equiv \ \langle \rho \rangle \ = \ \frac{3\rho_{\rm cr}}{\pi^2} \, ,
\end{align}
which will be used in~\eqref{eq:dng}. 

To calculate the change in the speed of the GWs due to the change in refractive index~\eqref{eq:dng}, let us assume that the GW is produced at a distance $D$ from Earth and encounters a spherical BEC-DM halo of radius $R$ en route.\footnote{For multiple DM halos along the line of slight, the calculation presented here can be extended in a straightforward manner, since it is only sensitive to the {\it total} distance traversed by the GWs through the DM halo(s).} The average fraction of distance the GW propagates through the DM halo with a reduced speed $c_g=1/n_g$ is given by
\begin{align}
x \ \equiv \ \frac{\langle D_{\rm halo}\rangle}{D} \ = \ \frac{4R}{\pi D}\, . 
\label{eq:x}
\end{align}
The effective speed of GWs is then given by 
\begin{align}
c_\text{eff} \ \equiv  \ \frac{D}{\Delta \tau} \ = \ \frac{c_g}{x + (1-x)c_g} \, ,
\end{align}
where $\Delta \tau = x D/c_g + (1-x) D$ is the proper time elapsed between the emission and detection of the GW signal. So the change in the speed of GWs from the speed of light in vacuum due to its encounter with the BEC-DM halo is given by 
\begin{align}
\delta c_g \ \equiv \ 1-c_{\rm eff} \ = \ \frac{x \delta n_g}{1+x \delta n_g} \, , 
\label{eq:deltacg}
\end{align}
where $\delta n_g$ is given by~\eqref{eq:dng}. This is our key result that will be used to put new constraints on the BEC-DM properties.

\section{GW Constraints on BEC-DM} \label{sec:3}

Using~\eqref{eq:deltacg}, we numerically calculate the change in the speed of GWs $\delta c_g$ as a function of the microscopic BEC-DM parameters $m$ and $\lambda$ for given values of the source distance $D$, GW frequency $f$ and amplitude $h$. For illustration, we will fix the central core density at $\rho_{\rm cr}=0.04 M_{\odot}/{\rm pc}^3$ (where $M_\odot$ is the solar mass), which is within the range suggested by a recent $N$-body simulation of self-interacting DM~\cite{Rocha:2012jg}. We also take $D=400$ Mpc, $f=35$ Hz and $h = 10^{-21}$ as representative values from the GW150914 event at LIGO~\cite{Abbott:2016blz}. The DM particle mass is varied in the range $m\in [10^{-23},1]$ eV. The upper limit comes from the basic condition that the particle's de Broglie wavelength, $\lambda_{\rm dB}=2\pi/mv$ (where $v\sim 10^{-3}$ is the virial velocity and we set $\hbar = 1$) should be larger than the inter-particle spacing, $d=(m/\rho)^{1/3}$, such that the wave functions of the individual particles in the system overlap with each other to form a BEC. The de Broglie wavelength also sets a natural lower limit to the core size of equilibrium BEC-DM halos that can form; taking $\lambda_{\rm dB}\lesssim 1$ kpc, the halo size of a typical dwarf spheroidal (DSph) galaxy, we get a lower limit of $m\gtrsim 5\times 10^{-23}$ eV, which is saturated for fuzzy DM~\cite{Hu:2000ke}. 

With this choice of parameters, we find the minimum $\delta c_g$ that can rule out repulsive BEC-DM is at the level of $10^{-37}$ for macroscopic GW parameters which LIGO is sensitive to. Thus we need the experimental sensitivity of $\delta c_g^{\rm exp}$ at this level or below to be able to put constraints on the BEC-DM parameter space using our method. For comparison, the current best model-independent bound is $\delta c_g\leq 2\times 10^{-15}$~\cite{Moore:2001bv}, deduced from the absence of gravitational Cherenkov radiation allowing for the unimpeded propagation of high-energy cosmic rays across our galaxy. Recently, assuming that the short gamma-ray burst above 50 keV detected by Fermi-GBM~\cite{Connaughton:2016umz} just 0.4 seconds after the detection of GW150914 at LIGO~\cite{Abbott:2016blz} originated from the same location, more stringent limits on $\delta c_g$ have been derived~\cite{Li:2016iww, Ellis:2016rrr, Collett:2016dey, Branchina:2016gad}. While a typical time-of-flight analysis~\cite{Nishizawa:2014zna} gives $\delta c_g \lesssim 10^{-17}$~\cite{Li:2016iww, Ellis:2016rrr, Collett:2016dey}, using modified energy dispersion relations (typical of many quantum gravity models) with the quantum gravity scale $E_G\geq M_{\rm Pl}$ yields a much stronger limit of $\delta c_g\lesssim 10^{-40}$~\cite{Branchina:2016gad}. However, whether the Fermi-GBM event originates from the same astrophysical source responsible for GW150914 is a controversial issue~\cite{Loeb:2016fzn, Lyutikov:2016mgv, Janiuk:2016qpe, Zhang:2016kyq, Woosley:2016nnw, Fraschetti:2016bpm, Greiner:2016dsk} and according to a recent analysis~\cite{Greiner:2016dsk}, the GBM event is more likely a background fluctuation, which is consistent with the non-detection of similar gamma-ray events at SWIFT~\cite{Evans:2016mta}, INTEGRAL~\cite{Savchenko:2016kiv}, and AGILE~\cite{Tavani:2016jrd}. Nevertheless, after the detection of the second LIGO event GW151226~\cite{Abbott:2016nmj}, the multi-messenger searches have become more intense and now include searches for gamma-ray~\cite{Racusin:2016fko, Adriani:2016gdx, Vianello:2016jzm}, $X$-ray~\cite{Evans:2016dgd, Adriani:2016gdx}, optical~\cite{Cowperthwaite:2016shk, Smartt:2016oeu} and neutrino~\cite{Adrian-Martinez:2016xgn, Gando:2016zhq, Aab:2016ras, Abe:2016jwn} counterparts. With more GW events expected from LIGO in the near future, these multi-messenger searches are likely to detect events coming from the same source and improve the limits on $\delta c_g$ significantly.  

Since the change in refractive index in a BEC medium is inversely proportional to the fourth power of the GW frequency [cf.~\eqref{eq:dng}], a future space-based GW interferometer, such as eLISA~\cite{Seoane:2013qna} with a lower operational frequency range of $0.1$--100 mHz, can further improve the sensitivity. For instance, for $f=1$ mHz, $D=3$ Gpc, $h=10^{-20}$ and $\rho_{\rm cr}$ same as above, the minimum $\delta c_g$ required to rule out BEC-DM is at the level of $10^{-24}$. Pulsar timing arrays, such as the ones united under IPTA~\cite{IPTA:2013lea} and SKA~\cite{Janssen:2014dka}, probe much lower frequencies around 1 nHz and amplitudes at the level of $h=10^{-14}$ and are capable of detecting $\delta c_g\sim 10^{-23}$, similar to the eLISA sensitivity and within reach of not-too-far-distant multi-messenger searches. 
\begin{figure*}[t!]
\centering
\includegraphics[width=8.5cm]{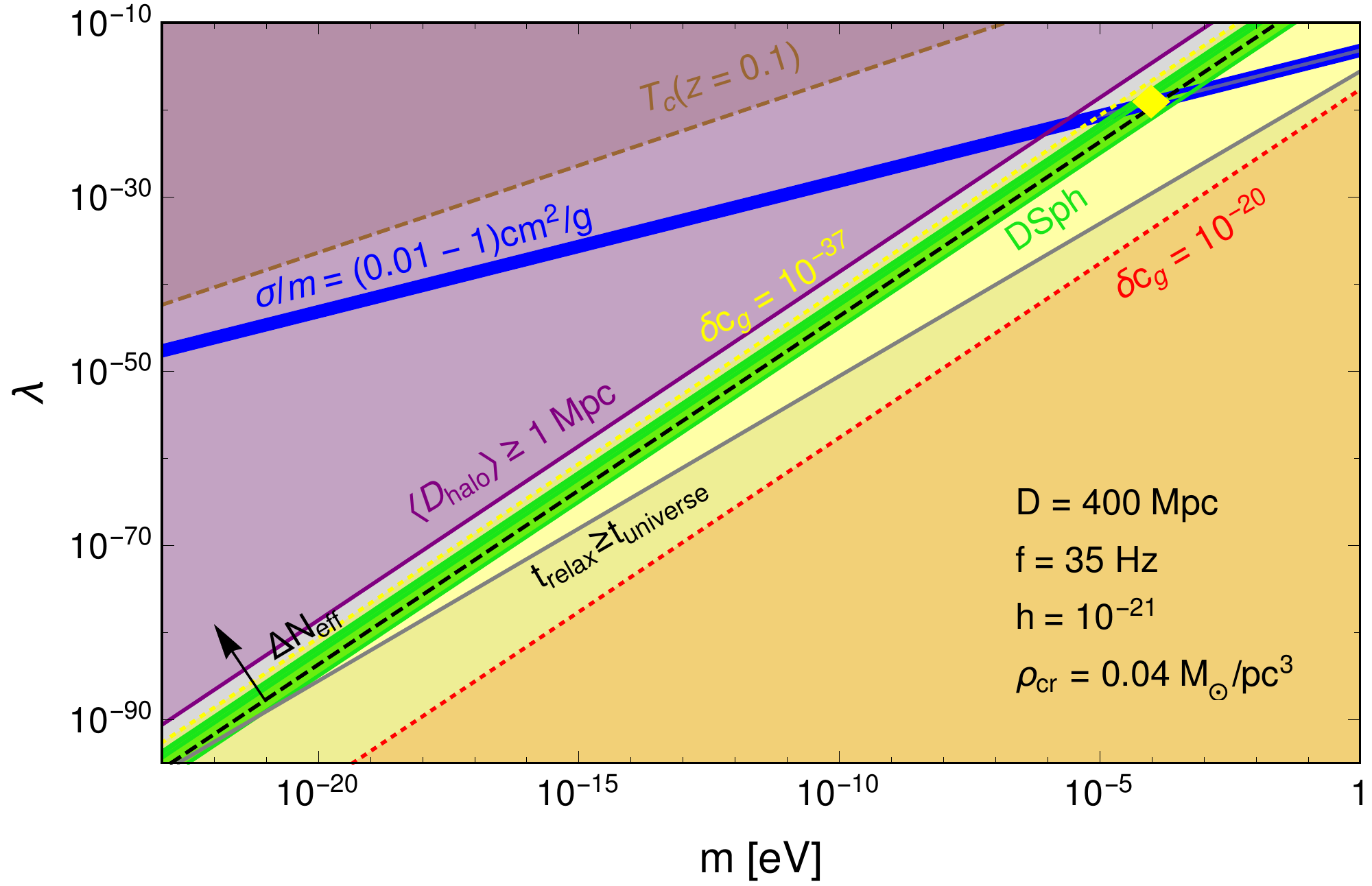}
\includegraphics[width=8.5cm]{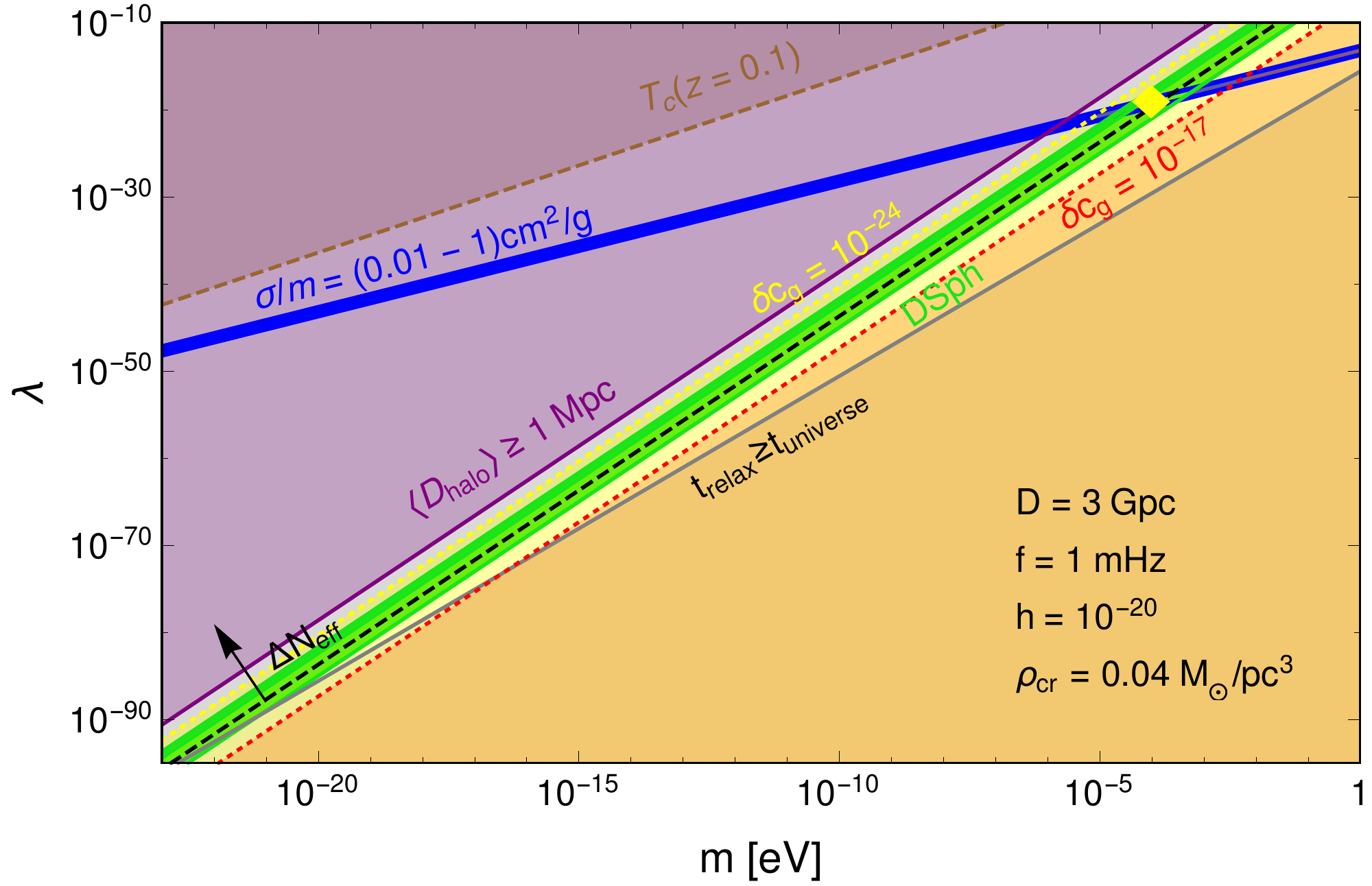}
\caption{{\it Left Panel:} Constraints on the quartic self-interaction as a function of the scalar mass for an intervening BEC-DM halo along the line of sight of a GW signal, where the red and orange shaded regions can be excluded for an upper bound on $\delta c_g$ of $10^{-20}$ and $10^{-37}$ respectively. Here we have used $f=35$ Hz and $h=10^{-21}$ for the GW frequency and amplitude respectively, and $D=400$ Mpc for the source distance as representative values from the LIGO event GW150914. We have also assumed the central core density of the BEC-DM halo $\rho_{\rm cr}=0.04 M_{\odot}/{\rm pc}^3$. {\it Right Panel:} Similar constraints from a representative future event at eLISA. For details of other constraints shown here, see text in Sections~\ref{sec:3} and \ref{sec:4}.  
%In both the plots, the blue band corresponding to $\sigma/m=$(0.01--1) cm$^2$/g is the preferred range for a self-interacting DM to solve the small-scale structure anomalies. The dark green band corresponds to a BEC-DM halo with radius $R=0.5$--10 kpc, as preferred by fits to DSph galaxy surveys. The overlap of the blue and green regions is physically favored, as shown by the yellow point which is completely within the exclusion region for $\delta c_g\leq 10^{-37}$ for LIGO and $\delta c_g\leq 10^{-24}$ for eLISA. The purple shaded region in the bottom half is excluded as the relaxation time to form a BEC exceeds the age of the universe. The black shaded region in the top left corner is excluded as the critical temperature falls below the temperature of the thermal plasma at the time of source redshift $z=0.1$, and above the dashed black line, it falls below the BBN temperature.  In the shaded region above the cyan solid line,  the average halo size $\langle D_{\rm halo}\rangle$ exceeds 1 Mpc. The dark blue dashed line gives a light deflection angle of $\delta \theta_{\rm def}=10^{-7}$ due to gravitational lensing by the intervening BEC-DM halo along the line of sight.
} 
\label{fig:1}
\end{figure*}

In Figure~\ref{fig:1} (left panel), we show the exclusion regions for $\delta c_g \leq 10^{-20}$  (red) and $\delta c_g \leq 10^{-37}$ (yellow) in the $(m,\lambda)$ plane using~\eqref{eq:deltacg} for a GW signal detected by LIGO with $D = 400$ Mpc, $f=35$ Hz and $h = 10^{-21}$. The dependence on the macroscopic GW parameters, and thus the possibility for a detection in future experiments, is demonstrated in Figure\ref{fig:1} (right panel) where the exclusion regions for $\delta c_g \leq 10^{-17}$ (red) and $\delta c_g \leq 10^{-24}$ (orange) for a hypothetical GW event detected by eLISA with $f=1$ mHz, $D=3$ Gpc, $h=10^{-20}$ are shown. For comparison, we also show the region which gives $\sigma/m=(0.01-1)~{\rm cm}^2/g$ (blue shaded), as preferred by $N$-body simulations to explain the small-scale structure anomalies, while being consistent with all observational constraints from colliding galaxy clusters~\cite{Markevitch:2003at, Randall:2007ph, Massey:2010nd, Kahlhoefer:2015vua,  Harvey:2015hha} and halo shapes~\cite{Vogelsberger:2012ku, Rocha:2012jg, Peter:2012jh, Zavala:2012us, Kaplinghat:2015aga}. Similarly, the viability of the BEC-DM halo model~\eqref{eq:halo} to fit the rotational curves of the most DM-dominated low surface brightness and DSph galaxies from different surveys implies $R\sim 0.5$--10 kpc~\cite{Boehmer:2007um, Arbey:2003sj, Harko:2011xw, Robles:2012uy, Diez-Tejedor:2014naa}, which can be translated to a preferred range of $m/\lambda^{1/4}\sim 4$--18 eV, as shown by the dark green shaded region in Figure~\ref{fig:1}. Note that the region of intersection between the blue and green shaded areas gives the physically preferred value of $(m,\lambda)\simeq (10^{-4}~{\rm eV}, 10^{-19})$, as shown by the yellow point. We find that for the LIGO frequency range $f=10$--350 Hz, the physical region can be completely excluded for $\delta c_g \leq 10^{-37}$. However, for the eLISA parameters the physically viable region can already be excluded for $\delta c_g \leq 10^{-24}$, which should be soon achievable in the multi-messenger approach. Further, eLISA will already be able to rule out complementary parameter space with current limits of $\delta c_g \leq 10^{-17}$.

The quantity $m^4/\lambda$ also gives a rough estimate of the total energy density of the DM field at the time of its transition from the radiation-like (when the scalar potential is dominated by the quartic term) to matter-like epoch (when the quadratic term in the scalar potential takes over the quartic term). The field density before this transition contributes to the extra relativistic species $\Delta N_{\rm eff}$ at Big Bang Nucleosynthesis (BBN)~\cite{Arbey:2003sj, Diez-Tejedor:2014naa, Li:2013nal}. Using the latest Planck result on $\Delta N_{\rm eff}\lesssim 0.39$~\cite{Ade:2015xua}, we obtain an additional constraint on $m/\lambda^{1/4}\gtrsim 8.5$ eV, as shown by the gray shaded region in Figure~\ref{fig:1}, which disfavors part of the DSph-preferred region. Future constraints on the extra relativistic degrees of freedom at BBN with the precision of $\Delta N_{\rm eff} \lesssim 0.12$ could rule out the entire DSph-preferred region of BEC-DM with repulsive self-interaction. 

Apart from the observational constraints, one should also satisfy important theoretical constraints from the BEC formation requirements. The orange shaded region in the lower right part of Figure~\ref{fig:1} is excluded as the relaxation time $t_{\rm relax}$ in the virialized DM clumps due to the scattering process $\phi\phi\to \phi \phi$ exceeds the age of the Universe $t_{\rm universe}$, which sets a lower limit on the self-interaction strength $\lambda \gtrsim 10^{-15}(m/{\rm eV})^{7/2}$ for the BEC to form~\cite{Tkachev:1991ka, Semikoz:1994zp, Riotto:2000kh}. Similarly, the brown shaded region in the top left part of the parameter space is disfavored, as the critical temperature $T_c=(24m^2/\lambda)^{1/2}$~\cite{Dolan:1973qd, Weinberg:1974hy, Kapusta:1981aa} below which a BEC can form, falls below the temperature of the universe at the source redshift of $z=0.1$ (assuming that the system of DM particles has a temperature comparable with that of radiation), which means the BEC-DM halo could not have formed at the time the GW was emitted from the binary black hole merger event GW150914. However, such a scenario would imply that there must exist extra relativistic components, in addition to the DM, to ensure thermal equilibrium, which contribute to $\Delta N_{\rm eff}$ at BBN and the corresponding constraint is much stronger than the $T(z=0.1)<T_c$ requirement, as can be seen from Figure~\ref{fig:1}.
%If we assume the DM particles were in thermal equilibrium with the SM particles through some additional interactions not shown in~\eqref{eq:lag}, this constraint becomes more severe, as in this case, $T_c$ is required to be larger than the BBN temperature $T_{\rm BBN}\sim 1$ MeV. 
Finally, an average halo size $\langle D_{\rm halo}\rangle \geq 1$ Mpc seems unrealistic for self-interacting DM halos and disfavored by simulations~\cite{Boehmer:2007um, Peter:2012jh}, as shown by the lighter purple shaded region in Figure~\ref{fig:1}. %In a more realistic scenario, the radius of the halo should be around 1 kpc, as shown by the green shaded region.

%%%%%%%%%%%%%%%%%%%%%%%
\section{Gravitational Lensing} \label{sec:4}
%%%%%%%%%%%%%%%%%%%%%%
In the multi-messenger approach, one way to confirm the existence of a BEC-DM halo in the path of the GW is by studying the deflection of photons passing through the region where galactic rotation curves are flat. The deflection angle is given by the standard formula~\cite{Misner:1974qy}
\begin{align}
\delta \theta_{\rm def} \ = \ \frac{4GM}{b} \, ,
\label{eq:def}
\end{align}
where $b$ is the impact parameter (i.e. distance of closest approach) for which we use the radius of the BEC-DM halo from~\eqref{eq:R2} and $M$ is the total mass of the DM halo, given by 
\begin{align}
M \ = \ 4\pi \int_0^R \rho(r)r^2 dr \ = \ \frac{4}{\pi}\rho_{\rm cr}R^3 \, , 
\label{eq:mass}
\end{align}
using~\eqref{eq:halo} for the density profile. \eqref{eq:def} is valid in the limit $GM\ll b$ which is satisfied in our case as long as $R\ll {\cal O}(1~{\rm Mpc})$. Thus, for a BEC-DM halo, we can express~\eqref{eq:def} completely in terms of the microscopic parameters:
\begin{align}
\delta \theta_{\rm def} \ = \ \frac{24\lambda}{m^4}\rho_{\rm cr}  \ =\ \frac{2R^2}{\pi^2 M^2_{\rm Pl}}\rho_{\rm cr} \, .
\label{eq:def2}
\end{align}
The physically interesting yellow point in Figure~\ref{fig:1} corresponds to a halo radius of $\sim 1$ kpc where we find $\delta \theta_{\rm def}=10^{-7}$. For other values of $R$, the prediction for the deflection angle can be readily obtained from~\eqref{eq:def2}. 
 
We should also clarify that the gravitational potential of the intervening DM halo along the line of sight will cause a Shapiro time delay~\cite{Shapiro:1964uw} for the GW, as well as its multi-messenger counterparts. In the geometrical optics approximation, treating the total gravitating mass as a point source, the time delay is the same for GW, photons and neutrinos, given by the general formula~\cite{Longo:1987gc, Krauss:1987me} 
\begin{align}
\Delta t_{\rm Shapiro} \ = \ (1+\gamma)GM \ln{\left(\frac{D}{b}\right)} \, ,
\label{eq:shapiro}
\end{align}
where $\gamma$ is a parametrized post-Newtonian parameter. However, this geometrical approximation breaks down for GWs with wavelengths larger than the size of the lensing object, which corresponds to lens masses $\lesssim 10^5M_\odot(f/{\rm Hz})^{-1}$. This can induce a differential Shapiro delay between the GW and photons/neutrinos of up to $0.1\:{\rm sec}(f/{\rm Hz})^{-1}$~\cite{Kahya:2016prx, Takahashi:2016jom}. For $R\sim 1$ kpc, we estimate the mass of BEC-DM halo from~\eqref{eq:mass} to be $M\sim 10^8M_\odot$; so for the LIGO frequency range of 10-350 Hz, the geometrical optics approximation~\eqref{eq:shapiro} remains valid and there is no relative Shapiro time delay to be considered in the multi-messenger analysis. However, for smaller frequencies, such as those relevant for eLISA and IPTA, the additional time delay must be taken into account while deriving experimental bounds on $\delta c_g$. 

%%%%%%%%%%%%%%%%%%%%%%%%%%%%%%%
\section{Conclusion} \label{sec:5}
%%%%%%%%%%%%%%%%%%%%%%%%%%%%%%
We have proposed a new method to probe BEC-DM using GW astronomy. We have shown that GWs passing through a BEC-DM halo will get appreciably slowed down due to energy loss in collective phononic excitations. The effective refractive index depends only on the mass and quartic coupling of the DM particles, apart from the frequency and amplitude of the propagating GW. Thus, an observable deviation $\delta c_g$ in the speed of GW can be used to put stringent constraints on the BEC-DM parameter space, as demonstrated in Figure~\ref{fig:1}. The physically interesting region of BEC-DM parameter space satisfying all existing constraints can be completely probed by this new method for $\delta c_g \leq 10^{-37}$ in the LIGO frequency range and $\delta c_g \leq 10^{-24}$ in the eLISA frequency range, which is soon achievable in a multi-messenger approach to GW astronomy. %Significantly improved sensitivity can be achieved at lower GW frequencies.

\acknowledgments
%\section*{Acknowledgments} 
This work of B.D. was supported in part by the DFG grant RO 2516/5-1. We thank the participants of the PhD Student Seminar on Gravitational Waves at MPIK %(\url{https://www.mpi-hd.mpg.de/personalhomes/smirnovj/Seminar/PhD\%20Seminar2016.html}) 
for pedagogical talks and useful discussions. B.D. also acknowledges helpful discussions with Nick Mavromatos and the local hospitality at ECT*, Trento where part of this work was done.

%%%%%%%%%%%%%%%%%%%%%%%%%%%%%

%%%%%%%%%%%%%%%%%%%%

\end{document}